


\magnification=\magstep1
\font\sectionfont=cmbx12 scaled\magstep1
\font\titlefont=cmbx10 scaled\magstep1

\def\AJ{AJ}
\def\AandA{A\&A}
\def\ApJ{ApJ}
\def\ApJS{ApJS}
\def\PRC{Phys.\ Rev.\ C}
\def\PRD{Phys.\ Rev.\ D}
\def\PRL{Phys.\ Rev.\ Lett.}

\def\section#1{{\bigbreak\leftline{\sectionfont #1}\nobreak\medskip}}

\newcount\eqnumber\eqnumber=0
\def\eqn{\global\advance\eqnumber by 1(\number\eqnumber)}

\def\hy#1{{${}^{#1}$H}}
\def\he#1{{${}^{#1}$He}}
\def\li#1{{${}^{#1}$Li}}
\def\be#1{{${}^{#1}$Be}}
\def\bo1#1{{${}^{1#1}$B}}
\def\ca1#1{{${}^{1#1}$C}}
\def\refitem{\par\noindent\hangindent=30pt\hangafter=1}


\baselineskip=20pt
    {
      \baselineskip=12pt
      \rightline{astro-ph/9412039}
      \rightline{OSU-TA-25/94}
      \rightline{December, 1994}
    }
  \vskip 3 cm
    \centerline{\titlefont Monte-Carlo Analysis of Big Bang Production}
    \centerline{\titlefont of Beryllium and Boron}
    \vskip .5cm
    \centerline{David Thomas$^{\dag}$}
    \vskip .3cm
    \centerline{Department of Physics, The Ohio State University}
    \centerline{Columbus, OH 43210}
  \vskip .5cm
    \centerline{\bf Abstract}
    {\noindent\narrower
    There is continued interest in the possibility that big bang
    nucleosynthesis
    may produce significant quantities of Be and B.  In this paper we
    reevaluate the primordial abundances taking into account uncertainties
    in reactions rates. We discuss the implications for primordial
    nucleosynthesis, and for galactic cosmic ray spallation.\par}
  \vskip 1cm
  \noindent{\it Submitted to The Astrophysical Journal}
  \vfill
  \hrule width 2 truein
  \noindent$^{\dag}$ thomas@pacific.mps.ohio-state.edu

\vfil\eject

\section{1 Introduction}
Big bang nucleosynthesis (BBN) has, for almost three decades, been the
standard model for the production of the light elements (\hy1, and \he4,
with trace amounts of \hy2, \he3 and \li7, see Walker et al., 1991; Copi,
Schramm and Turner; 1994, Thomas et al.\ 1995).  It is the absence of
stable isotopes at masses five and eight which prevents the build-up of
significant quantities of heavier species in the big bang.  It is possible,
however, that some small leakage may occur past this bottleneck, especially
in models (such as baryon-inhomogeneous models) which contain regions with
high neutron-to-proton ratios.  Indeed, as was pointed out by Boyd and
Kajino (1989) and by Malaney and Fowler (1989), detection of a primordial
abundance of an element like \be9 could provide the much-needed test by
which we may determine if there were significant inhomogeneities in baryon
number in the early universe.

Recently, Thomas et al.\ (1993, hereafter TSOF) have calculated primordial
abundances in the standard (homogeneous) model
using a much expanded reaction network, with the aim of
determining abundances of \li6, \be9 and B.  Their conclusions were that in
spite of some uncertainties in the reaction rates leading to \be9, it was
unlikely that BBN would be able to produce Be or B in the quantities
necessary to make detection possible.  Furthermore, it has been shown
(Thomas et al., 1994) that inhomogeneous models are unable to produce any
more Be or B (than the standard, homogeneous, model) without violating the
observed limits on the abundances of the light elements.  It seems then,
that not only is primordial \be9 unlikely to provide the desired litmus
test between the two models, but it may not be observable under any model
of BBN.

The current work on homogeneous production of Be and B
is motivated by two recent developments.  Data on \be9 in
old, population II, halo dwarfs has been improving rapidly in the last few
years.  Indeed Boesgaard and King (1993) have suggested that we may be
seeing the first sign of a plateau in the \be9 vs.\ metallicity data.  Such
a plateau would suggest that the primordial abundance had been discovered.
In addition, TSOF concerned themselves largely with uncertainties in the
rates of reactions between Be and B and other heavier elements.  While
these reactions are certainly important, reactions among the light elements
can also have a large effect on the abundances of LiBeB, even though the
uncertainties are smaller.  Since all species are produced from an initial
state consisting only of protons and neutrons, by transforming these
through the light elements into the heavier ones, all species are sensitive
to rates of reactions between the light elements.  In this paper, we
reevaluate the abundances of \li6, \be9 and B using a monte-carlo technique
to take into account the uncertainties in these reactions, and give
2-$\sigma$ uncertainties in the abundances.

\section{2 Observations}
The observational data on Be and B in metal-poor halo stars
is summarized in figures 1 and 2.  The
Be data (Rebolo et al., 1988; Ryan et al., 1992; Gilmore, Edvardsson and
Nissen, 1991; Gilmore et al., 1992; Boesgaard and King, 1993) shows a
definite dependence on metallicity.  (We use the usual astrophysical
conventions,
$
[\hbox{X}]\equiv12.0+\log(\hbox{X}/\hbox{H})
$
and
$
[\hbox{X}/\hbox{H}]\equiv \log(X/H)-\log(X/H)_\odot
$
.) If the primordial abundance of \be9 were within the range covered by the
observations, one would expect to see a plateau as the metallicity goes
to zero---i.e.\ the same \be9
abundance over a range of metallicities.  Figure 1 shows little, if any,
evidence of a plateau, suggesting that the primordial value is below
$[\hbox{Be}]=-1$.  We can take $[\hbox{Be}]\geq-2$ to be a (very) safe lower
limit on the observations.

The data on B is less abundant (Duncan, Lambert and Lemke, 1992; Edvardsson
et al., 1994), however there again appears to be an increase in abundance
with metallicity.  We can safely take $[\hbox{B}]\geq-1$ as a lower limit on
the observations.

There are currently only two observations of \li6 in metal-poor halo stars
(Smith, Lambert \& Nissen, 1993; Hobbs \& Thorburn, 1994).  These show \li6
at a few percent of total Li (\li6/H $\sim\hbox{a few}\times10^{-12}$).

\section{3 Results}
The theoretical predictions comes from a monte-carlo calculation generating
1000 data points for each value of $\eta$.  The code is based on the original
code of Wagoner (1967; Wagoner, Fowler \& Hoyle, 1969).  The reaction
network has been updated since then; our network is based on that in
Thomas et al.\ (1994).  Smith et al.\ (1993) have summarized the uncertainties
in the light element reaction rates.  We adopt their
values, with the exception that we use the latest world average for the
neutron lifetime (Particle Data Group, 1994), $\tau_n = 887.0\pm2.0$.
The calculation is for standard BBN---homogeneous, with three light neutrino
species.

Figure 3 shows primordial abundances (number densities relative to
hydrogen) for \li6, \be9, \bo10, \bo11 and for the sum of all species with
$A\ge12$.  For each species two-sigma bounds are shown.  The dip in the
\bo11 curve is due to the fact that it can be produced directly (for low
$\eta$) or as \ca11 (for high $\eta$) which then $\beta$-decays to \bo11.
It is worth noting that the uncertainty in \be9 and \bo10 is approximately
the same as that due to the uncertainty in the rate of \li7(t,n)\be9
(TSOF), although we note that this rate has been measured
(Brune et al., 1991).

The best-fit value of $\eta$ is currently the subject of some debate, and
will be discussed in detail in a future paper (Thomas et al., 1995).  For
further detail see, for example, Walker et al.\ (1991), Copi, Schramm and
Turner (1994), Krauss and Kernan (1994).  Here, we will simply defer any
discussion and assume that $\eta_{10} \equiv 10^{10}\times\eta$ is
somewhere between 1 and 10, with a best value around 3 (Walker et
al.\ 1991).  The relevant abundances are then those shown (with two-sigma
errors) in table 1.

\midinsert
\centerline{\bf TABLE 1}
\centerline{\bf Abundances}
\halign to \hsize{%
  \hfil#\tabskip=2em plus 1em&\hfil#&\hfil#&\hfil#\tabskip=0em\cr
  \noalign{\smallskip\hrule\smallskip\hrule\smallskip}
    &$\eta_{10}=1\hfil$&$\eta_{10}=3\hfil$&$\eta_{10}=10\hfil$\cr
  \noalign{\smallskip\hrule\smallskip}
    \li6/H & $(1.6\pm0.2)\times10^{-13}$ & $(3.2\pm0.5)\times10^{-14}$ &
             $(5.0\pm1.0)\times10^{-15}$ \cr
    \be9/H & $(8.5\pm6.0)\times10^{-17}$ & $(1.6\pm0.9)\times10^{-18}$ &
             $(4.0\pm2.0)\times10^{-20}$ \cr
    \bo10/H& $(2.2\pm1.0)\times10^{-19}$ & $(1.2\pm0.1)\times10^{-19}$ &
             $(8.0\pm1.5)\times10^{-21}$ \cr
    \bo11/H& $(6.6\pm3.6)\times10^{-18}$ & $(1.3\pm1.0)\times10^{-17}$ &
             $(1.3\pm0.5)\times10^{-15}$ \cr
($A\geq12$)/H& $(6.1\pm3.0)\times10^{-15}$ & $(2.4\pm1.1)\times10^{-14}$ &
             $(6.3\pm2.8)\times10^{-14}$ \cr
  \noalign{\smallskip\hrule\smallskip}
  }
\endinsert

\be9 takes its largest value at $\eta_{10}=1$.  Here, it is at least two
orders of magnitude below the smallest observed abundance in figure 1.
The uncertainty in the predicted abundance is not enough to make up
this difference.  It is conceivable that \be9 might reach the observed
levels for sufficiently small $\eta$ (TSOF), but this would imply a
\he4 abundance much lower than that observed in extragalactic HII
regions.  In addition, the excess \li7 produced would be a problem for
models of halo stars, since the \li7 would have to be depleted to the
level of the observations found on the Spite plateau.  Meanwhile, at
such low $\eta$, \hy2 and \he3 would be driven up to a value which
would require a complete change in our understanding of galactic
chemical evolution.

The \bo10 abundance is consistently lower than that of \bo11.  Since
observations are sensitive only to total B abundance this allows us to
concentrate on \bo11.  At $\eta_{10}=3$, \bo11 is close to its minimum
value of $\sim10^{-18}$.  Its largest value ($\sim10^{-15}$, for
$\eta_{10}\sim10$) is still two orders of magnitude below the
observations.  Once again, the uncertainty in the predicted abundance
has no significant effect on this conclusion.  The abundance does
increase with $\eta$ (TSOF), but like the solution to \be9 above, this
causes difficulties reconciling the light element abundances with
observations.  High values of $\eta$ overproduce both \li7 and \he4.

The \li6 abundance at $\eta_{10}=3$ is a factor of 100 lower than the
observations.  We can reduce this discrepancy by going to a lower
value of $\eta$, but again this underproduces \he4 and overproduces
\li7.  Worse still, this scenario would require significant depletion
of \li7 in population II stars and this implies depletion of \li6
(Brown \& Schramm, 1988).  Thus, the \li6 discrepancy reappears.

It seems fairly conclusive then, that the primordial abundances of
\li6, \be9 and B are significantly lower than the current set of observations,
and are likely to remain out of reach for the foreseeable future.
Given the correlation between Be/B abundances and metallicity, this
comes as no great surprise.  It is likely that the observed abundances
are due to processes in the early galaxy, such as cosmic ray
spallation of the interstellar medium.  Cosmic rays are believed to be
responsible for the abundances of \li6, \be9, \bo10 and \bo11 in
population I stars (Walker, Mathews and Viola, 1985).  The same
mechanism acting on population II stars can produce significant
amounts of Li, while remaining consistent with the observed Be
abundances (Steigman and Walker, 1992) and with a minimal set of
assumptions it is possible to reproduce the observed abundances of
LiBeB (Walker et al., 1993).

\section{Acknowledgments}

I would like to thank Terry Walker for comments on the manuscript.
This work was supported in part by the Department of Energy grant
DE-AC02-76ER-01545.

\section{References}

\refitem Boesgaard, A.M., \& King, J.R. 1993, \AJ, {\bf 106}, 2309
\refitem Boyd, R.N., \& Kajino, T. 1989, \ApJ, {\bf 336}, L55
\refitem Brown, L., \& Schramm, D.N. 1988, \ApJ, {\bf 329}, L103
\refitem Brune, C.R., Kavanagh, R.W., Kellogg, S.E., \& Wang, T.R. 1991,
         \PRC, {\bf 43}, 875
\refitem Copi, C.J., Schramm, D.N., \& Turner, M.S. 1994,
         Science, (in press), FERMILAB-Pub-94/174-A
\refitem Duncan, D.K., Lambert, D.L., \& Lemke, M. 1992,
         \ApJ, {\bf 401}, 584
\refitem Edvardsson, B., Gustafsson, B., Johansson, S.G.,
         Kiselman, D., Lambert, D.L., Nissen, P.E., \& Glimore, G. 1994,
         \AandA, {\bf 290}, 176
\refitem Gilmore, G., Edvardsson, B., \& Nissen, P.E. 1991,
         \ApJ, {\bf 378}, 17
\refitem Gilmore, G., Gustafsson, B., Edvardsson, B., \& Nissen, P.E. 1992,
         Nature, {\bf 357}, 379
\refitem Hobbs, L., \& Thorburn, J.A. 1994, \ApJ, {\bf 428}, L25
\refitem Kernan, P., \& Krauss, L.M. 1994, \PRL, {\bf 72}, 3309
\refitem Krauss, L.M., \& Kernan, P. 1994, Case Western preprint
         CWRU-P9-94-REV
\refitem Krauss, L.M., \& Romanelli, P. 1990, \ApJ, 358, 47
\refitem Malaney, R.A., \& Fowler, W.A. 1989, \ApJ, {\bf 345}, L5
\refitem Particle Data Group 1994, \PRD, {\bf 50} 1173
\refitem Rebolo, R., Molaro, P., Abia, C., \& Beckman, J.E. 1988,
         \AandA, {\bf 193}, 193
\refitem Ryan, S.G., Norris, J.E., Bessell, M.S., \& Deliyannis, C.P. 1992,
         \ApJ, {\bf 388}, 184
\refitem Smith, M.S., Kawano, L.H., \& Malaney, R.A. 1993,
         \ApJS, {\bf 85}, 219
\refitem Smith, V.V., Lambert, D.L., \& Nissen, P.E. 1993,
         \ApJ, {\bf 408}, 262
\refitem Steigman, G., \& Walker, T.P. 1992, \ApJ, {\bf 385}, L13
\refitem Thomas, D., Schramm, D.N., Olive, K.A., \& Fields, B.D.
         1993, \ApJ, {\bf 406}, 569 (TSOF)
\refitem Thomas, D., Schramm, D.N., Olive, K.A., Mathews, G.J.,
         Meyer, B.S., \& Fields, B.D. 1994, \ApJ, {\bf 430}, 291
\refitem Thomas, D., Hata, N., Scherrer, R.J., Steigman, G., \& Walker, T.P.
         1995, in preparation
\refitem Wagoner, R.V. 1969, \ApJS, {\bf 18}, 247
\refitem Wagoner, R.V., Fowler, W.A., \& Hoyle, F. 1967, \ApJ, {\bf 148}, 3
\refitem Walker, T.P., Mathews, G.J., \& Viola, V.E. 1985,
         \ApJ, {\bf 299}, 745
\refitem Walker, T.P., Steigman, G., Schramm, D.N., Olive, K.A., \&
         Fields, B.D. 1993, \ApJ, {\bf 413}, 562
\refitem Walker, T.P., Steigman, G., Schramm, D.N., Olive, K.A., \&
         Kang, H. 1991, \ApJ, {\bf 376},  51

\vfil\eject
\section{Figure Captions}

\item{[1.]} Observations of Be in halo dwarfs, as a function of metallicity.
            Data is taken from Rebolo et al., 1988 (RMAB); Gilmore et
            al., 1991 (GEN); Gilmore et al., 1992 (GGEN); Ryan et al.,
            1992 (RNBD); Boesgaard \& King, 1993 (BK).
\item{[2.]} Observations of B  in halo dwarfs, as a function of metallicity.
            Data is taken from Duncan et al., 1992 (DLL); Edvardsson et
            al., 1994 (EGJ).
\item{[3.]} Big Bang Nucleosynthesis abundances (number densities,
            relative to hydrogen) of \li6, \be9, \bo10,
            \bo11 and the sum of all species with $A\geq12$.
            Dashed lines show 2-$\sigma$ bounds.

\bye